\documentclass[10pt, aps, pra, twocolumn, superscriptaddress, floatfix, showpacs, longbibliography]{revtex4-1}

\usepackage{graphicx}
\usepackage{amsfonts}
\usepackage{amssymb}
\usepackage{amsmath}
\usepackage{txfonts}
\usepackage{lipsum}
\usepackage{color}
\usepackage{wasysym}
\usepackage[colorlinks=true, allcolors=blue]{hyperref}
\usepackage{bbold}
\usepackage{etoolbox} 
\usepackage[normalem]{ulem}
\usepackage{mathtools} 
\usepackage{multirow} 
\usepackage{hhline}
\usepackage{mathrsfs} 
\usepackage{soul}

\usepackage{letltxmacro}
\LetLtxMacro{\oldsqrt}{\sqrt}
\renewcommand{\sqrt}[2][\mkern8mu]{\mkern-6mu\mathop{}\oldsqrt[#1]{#2}}

\usepackage{url}
\definecolor{indigo(dye)}{rgb}{0.0, 0.25, 0.42}
\usepackage{placeins}

\begin{document}

\title{Fingerprints of a charge ice state in the doped Mott insulator Nb$_3$Cl$_8$}

\author{Evgeny~A.~Stepanov}
\affiliation{CPHT, CNRS, {\'E}cole polytechnique, Institut Polytechnique de Paris, 91120 Palaiseau, France}
\affiliation{Coll\`ege de France, Universit\'e PSL, 11 place Marcelin Berthelot, 75005 Paris, France}

\begin{abstract}
The interplay between strong electronic correlations and the inherent frustration of certain lattice geometries is a common mechanism for the formation of nontrivial states of matter. In this work, we theoretically explore the collective electronic effects in the monolayer Nb$_3$Cl$_8$, a recently discovered triangular lattice Mott insulator. Our advanced many-body numerical simulations predict the emergence of a phase separation region upon doping this material. Notably, in close proximity to the phase separation, the static charge susceptibility undergoes a drastic change and reveals a distinctive bow-tie structure in momentum space. The appearance of such a fingerprint in the context of spin degrees of freedom would indicate the formation of a spin ice state. This finding allows us to associate the observed phase separation to a charge ice state, a state with a remarkable power law dependence of both the effective exchange interaction and correlations between electronic densities in real space.
\end{abstract}

\maketitle

\section{Introduction}

Mott insulators are among the most prominent examples of materials where the insulating behaviour does not arise from the band gap in the non-interacting electronic spectral function. 
Instead, it occurs as a result of strong local Coulomb repulsion that localizes the electrons on a lattice~\cite{Mott}.
These localized electrons typically form well-developed magnetic moments, leading to spatial magnetic fluctuations and the potential for a spin-ordered ground state at low temperatures. 
In this respect, Mott insulators are particularly promising for realizing exotic states such as the valence-bond, spin ice, or (quantum) spin liquid states~\cite{balents2010spin}.
Doping Mott insulators usually suppresses magnetic fluctuations, but on the other hand, it may give rise to nontrivial collective electronic instabilities in the charge channel.
For instance, it has been shown that doped Mott insulators can exhibit a first-order phase transition between either the Mott insulating or strongly correlated metallic phase and the weakly correlated metallic phase~\cite{PhysRevLett.67.259, PhysRevLett.75.4650, PhysRevLett.89.046401, PhysRevLett.104.226402, PhysRevB.84.075161, PhysRevB.100.085104, PhysRevB.102.205127, PhysRevB.105.125140, PhysRevLett.130.066401, PhysRevB.110.L121109}. 
This transition is associated with the appearance of a phase separation (PS) region, which is signaled by the divergence of the electronic compressibility or the charge susceptibility at zero momentum.
Supplementing strong electronic correlation with the frustration that originates from a specific lattice geometry may result in the formation of more exotic phases. 
Thus, Mott insulators on a triangular lattice can reveal signatures of the chiral spin, charge, or superconducting states~\cite{PhysRevLett.120.196402, vandelli2024doping, ming2023evidence}. 

Finding an experimental realization of a Mott insulator is challenging. 
It requires identifying a system with a half-filled narrow band, which, without considering electronic interactions, is located at the Fermi energy and is distinctly decoupled from other bands. 
Furthermore, the Hubbard bands emerging from this narrow band upon including the interaction should also not hybridize with the rest of the energy spectrum.
In this context, the recently discovered Mott insulator Nb$_3$Cl$_8$, with an effective triangular geometry arising from the distorted kagome lattice, appears to be a promising candidate for realizing nontrivial states of matter~\cite{hu2023correlated}.
According to {\it ab-initio} theoretical calculations, this material exhibits a narrow half-filled band at the Fermi level~\cite{doi:10.1021/acs.nanolett.2c00778, PhysRevX.13.041049, grytsiuk2024nb3cl8, PhysRevB.107.035126} and a very strong local interaction, which, in some model parametrizations, is an order of magnitude larger than the bandwidth of the non-interacting electronic dispersion~\cite{grytsiuk2024nb3cl8}. 
Such a strong Coulomb repulsion leads to a single-electron occupation of each lattice site, which makes the material a Mott insulator with a rather large gap.
It has been observed that at high temperatures, both the monolayer and bulk phases of Nb$_3$Cl$_8$ are paramagnetic, as confirmed by the Curie-Weiss behaviour of the spin susceptibility~\cite{doi:10.1021/acsnano.9b04392, PhysRevX.13.041049}.
At low temperatures, the bulk system undergoes a structural transition, leading to a nonmagnetic singlet ground state~\cite{C6QI00470A, doi:10.1021/acs.inorgchem.6b03028}. 
In turn, the monolayer Nb$_3$Cl$_8$ is expected to form a 120$^{\circ}$ antiferromagnetic (AFM) ground state, as theoretically predicted based on Heisenberg model calculations.~\cite{grytsiuk2024nb3cl8}.  

The characteristics of doped Nb$_3$Cl$_8$ have not been extensively addressed yet, with current research primarily focusing on its magnetic properties~\cite{doi:10.1021/acsnano.9b04392} and conductivity~\cite{Yoon_2020}.
In doped Mott insulators, the strong local correlations that lead to the formation of a Mott insulating state are anticipated to compete with significant spatial collective electronic fluctuations. 
This interplay may give rise to complex many-body effects.
The recent exciting results of Refs.~\onlinecite{PhysRevLett.120.196402, vandelli2024doping, PhysRevB.110.L121109}, obtained for other triangular lattice Mott insulators, strongly encourage us to investigate collective electronic instabilities that may arise in the monolayer Nb$_3$Cl$_8$ upon doping.

Our findings reveal that doping this material leads to phase separation driven by the formation of a charge ice state.
This state is evidenced by the distinct behavior of two independent quantities: the static charge susceptibility and the effective exchange interaction between charge densities, both exhibiting a remarkable power law dependence in real space. 
This dependence is reminiscent of hydrogen bonding interactions in water and is one of the distinct fingerprints of a spin ice state if it is found in the context of spin degrees of freedom.
Our study not only theoretically addresses the charge ice state but also provides compelling evidence that this unique phase can potentially be realized in a real material.

\section{Model and method} 
\begin{figure}[t!]
\includegraphics[width=0.8\linewidth]{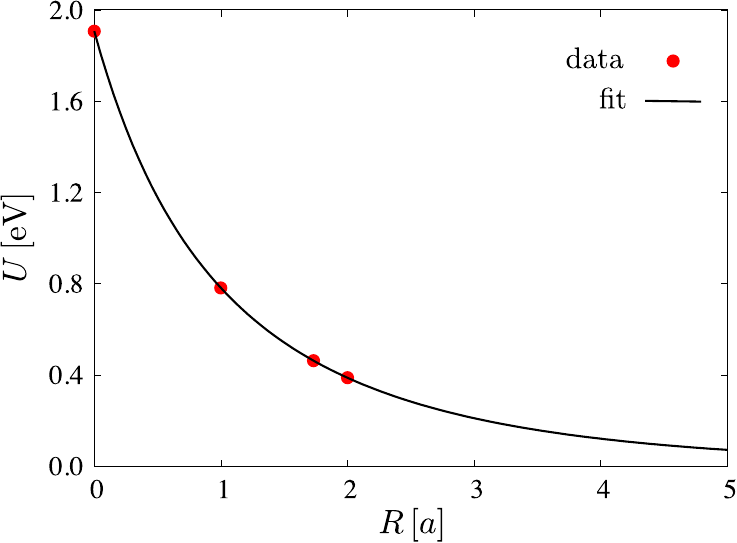}
\caption{Coulomb interaction. The red dots represent the values of the Coulomb interaction $U(R)$ up to the third nearest-neighbor distance in real space obtained in Ref.~\onlinecite{grytsiuk2024nb3cl8}. The solid black line corresponds to the Yukawa-like form of the fit function introduced to account for the long-range tail of the Coulomb interaction. 
\label{fig:U}}
\end{figure}

To investigate the many-body effects in the monolayer Nb$_3$Cl$_8$, we use a single molecular orbital extended Hubbard model on an effective triangular lattice that was introduced in Ref.~\onlinecite{grytsiuk2024nb3cl8} based on {\it ab-initio} calculations.
The corresponding model Hamiltonian
\begin{align}
H = -\sum_{j,j',\sigma} t^{\phantom{\dagger}}_{jj'} c^{\dagger}_{j\sigma} c^{\phantom{\dagger}}_{j'\sigma} 
- \mu \sum_{j}n_j
+ \frac12 \sum_{j,j'} U_{|j\text{-}j'|} n_{j}n_{j'}
\label{eq:Hlatt}
\end{align}
describes the hopping of electrons between the lattice sites $j$ and $j'$ with the hopping amplitude $t_{jj'}$ by means of the operators $c^{(\dagger)}_{j\sigma}$ that annihilate (create) an electron on the site $j$ with the spin projection ${\sigma\in\{\uparrow,\downarrow\}}$.
The hopping amplitudes up to the third nearest-neighbor on a triangular lattice are: ${t_1=22.6\,\text{meV}}$, ${t_2=4.6\,\text{meV}}$ and ${t_3=-4.0\,\text{meV}}$. 
The chemical potential $\mu$ is included to the model to control the occupation of the orbital.
In Nb$_3$Cl$_8$, the Coulomb repulsion $U_{|j\text{-}j'|}$ between the electronic densities ${n_{j}=\sum_{\sigma}c^{\dagger}_{j\sigma}c^{\phantom{\dagger}}_{j\sigma}}$ is found to be extremely large compared to the non-interacting bandwidth and rather long-ranged~\cite{grytsiuk2024nb3cl8}: 
The local interaction is equal to ${U_{0}\simeq1.9}$\,eV and the interaction between the neighboring lattice sites $U_{1}$ is only approximately 2.5 times smaller than the local one.
The corresponding values of the Coulomb interaction $U_j$ are depicted by the red dots in Figure~\ref{fig:U}.
To reproduce the long-range tail of the Coulomb interaction we fit the real-space data $U(R)$ of Ref.~\onlinecite{grytsiuk2024nb3cl8} by the Yukawa-like potential ${f(R) = C_1\exp\{-C_2{}R\}/(R+C_3)}$, where ${C_1=2.67}$, ${C_2=0.35}$, and ${C_3=1.40}$.
We find that Yukawa potential (black curve in Fig.~\ref{fig:U}), which corresponds to a screened potential, better reproduces the short-range part of the Coulomb interaction than the conventional ${\sim1/R}$ form.

The undoped monolayer Nb$_3$Cl$_8$ resides in the Mott insulating state. 
An accurate description of this state, as well as other manifestations of strong collective electronic behavior, requires the use of advanced computational techniques. 
In this work, we employ the dual triply irreducible local expansion (D-TRILEX) approach~\cite{PhysRevB.100.205115, PhysRevB.103.245123, 10.21468/SciPostPhys.13.2.036}, which is ideally suited for this purpose.
This method consistently accounts for a combined effect of local correlations and spatial collective electronic fluctuations.

In \mbox{D-TRILEX}, the local correlations are taken into account non-perturbatively via the impurity problem of dynamical mean-field theory (DMFT)~\cite{RevModPhys.68.13}. 
The impurity problem is solved numerically exactly using the \textsc{w2dynamics} package~\cite{WALLERBERGER2019388} and provides the electron Green's function ${g_{\nu} = -\langle c^{\phantom{*}}_{\nu} c^{*}_{\nu} \rangle}$, the charge (${\varsigma=\text{ch}}$) and spin (${\varsigma=\text{sp}\in\{x,y,z\}}$) susceptibilities ${\chi^{\varsigma}_{\omega} = - \langle \rho^{\varsigma}_{\omega} \rho^{\varsigma}_{-\omega}\rangle}$, and the three-point vertex functions $\Lambda^{\varsigma}_{\nu\omega}$ that couple single-electron degrees of freedom to collective charge and spin fluctuations.
The latter are described by composite variables ${\rho^{\varsigma}_{\omega}=n^{\varsigma}_{\omega} - \langle n^{\varsigma}_{\omega}\rangle}$, where ${n^{\varsigma}_{\omega}=\sum_{\nu,\sigma\sigma'}c^{*}_{\nu+\omega,\sigma}\sigma^{\varsigma}_{\sigma\sigma'}c^{\phantom{*}}_{\nu\sigma'}}$ is the charge or spin density written in terms of fermionic Grassmann variables $c^{(*)}$, ${\sigma^{\rm ch}=\mathbb{1}}$ is the unitary matrix in the ${2\times2}$ spin space, and ${\sigma^{x,y,z}}$ are the Pauli matrices in the same space. $\nu$ and $\omega$ denote the fermionic and bosonic Matsubara frequencies, respectively.

The non-local correlations are incorporated beyond DMFT~\cite{RevModPhys.90.025003, Lyakhova_review} by considering the leading, i.e. particle-hole ladder, subset of Feynman diagrams describing electron scattering on collective charge and spin fluctuations.
The impurity quantities described above are used to construct the building blocks for the $GW$-like diagrammatic expansion, which is written in terms of the fermionic propagator $\tilde{G}_{{\bf{k}\nu}}$, the renormalized interaction $\tilde{W}^{\varsigma}_{{\bf{q}\omega}}$, and the impurity vertex function $\Lambda^{\varsigma}_{\nu\omega}$.
The diagrammatic expansion is performed using effective dual fermionic and bosonic variables, which enables the combination of both weak- and strong-coupling diagrammatic expansions within a unified framework.
Additionally, the transformation to dual space avoids double counting of correlations between the DMFT and the diagrammatic parts of the calculation.
Consequently, the method provides accurate results for the single-particle (electronic Green's function) and two-particle (charge and spin susceptibilities) response functions across a broad range of model parameters around both the weak- and strong-coupling limits~\cite{PhysRevB.103.245123, 10.21468/SciPostPhys.13.2.036}.

The diagrammatic part of \mbox{D-TRILEX} is computed self-consistently between single- and two-particle quantities based on the numerical implementation described in Ref.~\onlinecite{10.21468/SciPostPhys.13.2.036}.
At the end of the calculation, the exact transformation from the dual space to physical quantities is performed in order to obtain the electronic Green's function $G_{{\bf k}\nu}$, self-energy $\Sigma_{{\bf k}\nu}$, charge and spin susceptibilities $X^{\varsigma}_{{\bf q}\omega}$, and polarization operators $\Pi^{\varsigma}_{{\bf q}\omega}$.
These quantities are related through usual Dyson-like equations ${G^{-1}_{{\bf k}\nu} = G^{-1}_{0,{\bf k}\nu} - \Sigma^{\phantom{1}}_{{\bf k}\nu}}$ and ${X^{\varsigma\,-1}_{{\bf q}\omega} = \Pi^{\varsigma\,-1}_{{\bf q}\omega} - U^{\varsigma}_{{\bf q}}}$.
In these expressions, ${G^{-1}_{0,{\bf k}\nu} = i\nu + \mu - t_{\bf k}}$ is the inverse of the bare Green's function with $t_{\bf k}$ being the electronic dispersion and $U^{\varsigma}_{{\bf q}}$ is the bare interaction in the charge or spin channel. 
The momentum-resolved electronic spectral functions ${A({\bf k},E)}$ shown in this work are obtained from the Matsubara Green's functions $G_{{\bf k}\nu}$ via analytical continuation using the maximum entropy method implemented in the \textsc{ana\_cont} package~\cite{kaufmann2021anacont}. 
The static charge susceptibility $X^{\rm ch}_{\omega=0}({\bf q})$ discussed below is calculated at the zeroth bosonic Matsubara frequency ${\omega=0}$.

The applicability of \mbox{D-TRILEX} spans from model calculations~\cite{PhysRevLett.127.207205, PhysRevLett.129.096404, PhysRevResearch.5.L022016, PhysRevLett.132.236504, PhysRevLett.132.226501, PhysRevB.110.L161106} to realistic materials computations~\cite{stepanov2021coexisting, vandelli2024doping, stepanov2023charge, Ruthenates, Cuprates}. 
The method is particularly useful for detecting various ordered phases that are determined by the divergence of the corresponding susceptibilities at the momentum that defines the wave vector of the ordering. 
The strength of the fluctuations driving the phase transition can be estimated by looking at the largest static (${\omega=0}$) dielectric function $\epsilon^{\varsigma}_{\omega}({\bf q})$ in the corresponding (charge or spin) channel.
The dielectric function is related to the susceptibility as ${\epsilon^{\varsigma}_{\omega}({\bf q}) = \Pi^{\varsigma}_{\omega}({\bf q})/X^{\varsigma}_{\omega}({\bf q})}$ and shows how the polarization operator $\Pi^{\rm ch/sp}_{\omega}({\bf q})$ (irreducible with respect to the bare interaction ${U^{\rm ch/sp}({\bf q})}$ part of the susceptibility) is renormalized by the collective electronic fluctuations in the corresponding channel.
Thus, ${\epsilon=1}$ indicates the absence of the fluctuations, and ${\epsilon({\bf Q})=0}$ signals the formation of the ordered state with the ordering vector ${\bf q=Q}$.

\section{Results}

\subsection{Many-body effects at half filling}
We begin investigating many-body effects in the monolayer Nb$_3$Cl$_8$ with the half-filled case.
We perform calculations at two different temperatures, ${T=290\,\text{K}}$ and 145\,K.
At both temperatures, in agreement with the results of previous works~\cite{grytsiuk2024nb3cl8, doi:10.1021/acs.nanolett.2c00778, PhysRevX.13.041049, PhysRevB.107.035126}, we confirm that the material lies very deep in the Mott insulating phase: The obtained electronic spectral function features two narrow and nearly dispersiveless Hubbard bands separated by a large gap of the order of $U_{0}$.
However, we do not observe any signature of notable magnetic or charge fluctuations. 
Indeed, the largest dielectric function in the spin channel ${\epsilon^{\rm sp}_{\omega=0}({\bf q})}$ is found at the wave vector ${{\bf q}=\text{K}=\{4\pi/3, 0\}}$, which corresponds to the 120$^{\circ}$ AFM type of spin fluctuations, in agreement with the finding of Ref.~\onlinecite{grytsiuk2024nb3cl8}. 
However, by lowering the temperature from ${T=290}$\,K to 145\,K the spin dielectric function changes from ${\epsilon^{\rm sp}_{\omega=0}(\text{K})=0.97}$ to ${\epsilon^{\rm sp}_{\omega=0}(\text{K})=0.94}$, which indicates that the magnetic fluctuations are relatively weak.
Furthermore, a large extrapolated value of ${\epsilon^{\rm sp}_{\omega=0}(\text{K})=0.91}$ at ${T=0}$ indicates that the system does not exhibit any tendency toward the formation of the spin ordered state, at least at the considered temperatures.
Our calculations also do not reveal significant charge fluctuations.
The largest dielectric function in the charge channel corresponds to the zero momentum ${{\bf q}=\Gamma=\{0,0\}}$ and is very close to unity as well: ${\epsilon^{\rm ch}_{\omega=0}(\Gamma)=0.96}$ at ${T=290}$\,K and ${\epsilon^{\rm ch}_{\omega=0}(\Gamma)=0.97}$ at $T=145$\,K. 

\begin{figure}[t!]
\includegraphics[width=1\linewidth]{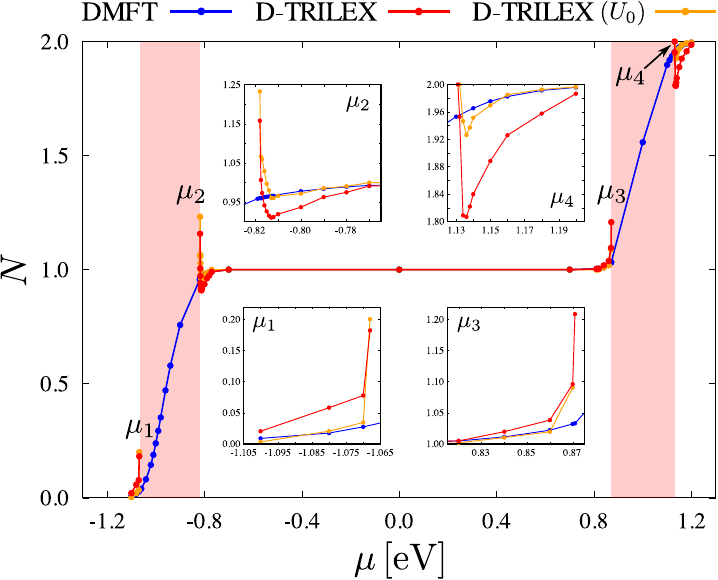}
\caption{Phase diagram. Dependence of the average electronic density $N$ on the value of the chemical potential $\mu$, where ${\mu=0}$ corresponds to the case of an undoped monolayer Nb$_3$Cl$_8$. Calculations are performed using DMFT with the local $U_0$ interaction (blue), \mbox{D-TRILEX} with the full long-range Coulomb potential $U_{|j\text{-}j'|}$ (red), and the ``\mbox{D-TRILEX} ($U_0$)'' method considering only the local interaction $U_0$ (orange). The chemical potentials $\mu_i$ correspond to the boundaries of the PS regions (shaded red areas). The result is obtained at ${T=290\,\text{K}}$. The insets show a more detailed behavior of the curves in the vicinity of PS regions.
\label{fig:Phase_diagram}}
\end{figure}

\begin{figure}[b!]
\includegraphics[width=1\linewidth]{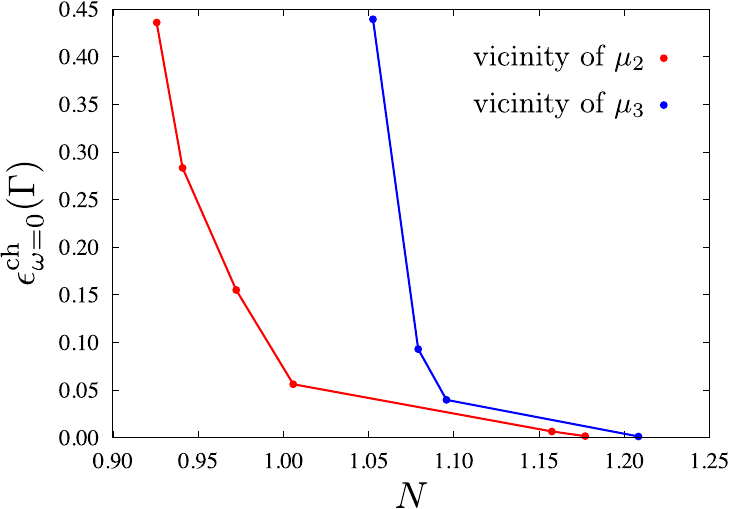}
\caption{Charge dielectric function in the vicinity of the PS. Evolution of the largest static dielectric function in the charge channel ${\epsilon^{\rm ch}_{\omega=0}(\Gamma)}$ as a function of the average electronic density $N$ calculated in the vicinity of the two PS boundaries defined by the chemical potentials $\mu_2$ (red) and $\mu_3$ (blue). The results are obtained at ${T=290\,\text{K}}$ using \mbox{D-TRILEX} with the full long-range Coulomb potential.
\label{fig:LE}}
\end{figure}

\begin{figure*}[t!]
\includegraphics[width=1\linewidth]{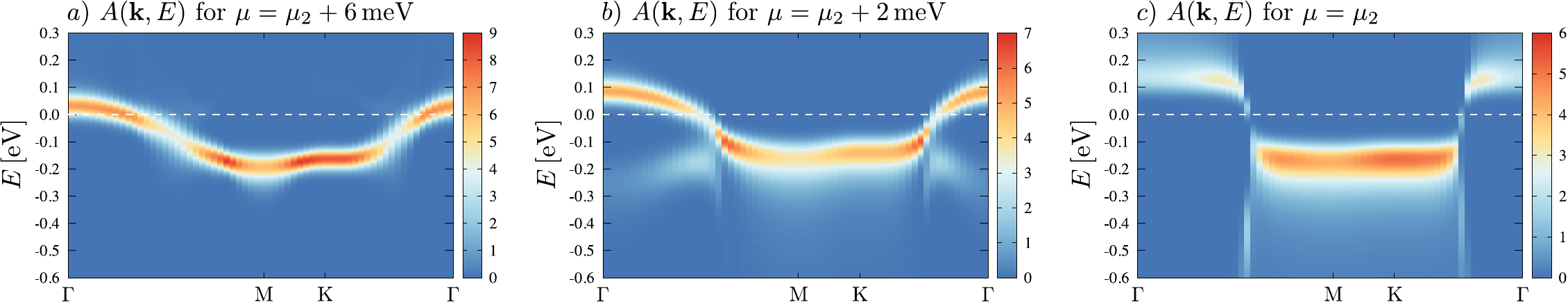}
\caption{Electronic spectral function. Evolution of the momentum-resolved electronic spectral function $A({\bf k},E)$ in the vicinity of the PS. Calculations are performed using \mbox{D-TRILEX} and take into account the effect of the full long-range Coulomb potential. The result is obtained at ${T=290\,\text{K}}$ for ${\mu=\mu_2+6\,\text{meV}}$ (a), ${\mu_2+2\,\text{meV}}$ (b) and ${\mu_2}$ (c), and plotted along the high-symmetry path in the BZ.
\label{fig:A_dop}}
\end{figure*}

\subsection{Phase separation and the charge ice state}

Doping the system does not alter the strength of the magnetic fluctuations, which are already weak in the considered material at half filling. 
However, we observe a significant change in the charge fluctuations upon doping.
Figure~\ref{fig:Phase_diagram} shows the evolution of the average electronic density $N$ as a function of the chemical potential calculated at ${T=290}$\,K.
Here, ${\mu=0}$ corresponds to the undoped case of a half-filled Mott insulator.
At ${|\mu|\lesssim0.8}$\,eV the chemical potential lies inside the gap and the system remains in the Mott insulating state with ${N=1}$.
At ${|\mu|\gtrsim0.8}$\,eV, taking into account only local correlations and the local Coulomb interaction $U_0$ within DMFT results in a gradual increase of the electronic density with increasing chemical potential (blue curve).
Importantly, no first-order phase transition is identified, which is consistent with other finite temperature single-site DMFT calculations of single-band models.
Additionally considering the non-local collective electronic fluctuations and the full long-range Coulomb potential $U_{|j\text{-}j'|}$ within the \mbox{D-TRILEX} approach drastically changes this picture (red curves). 
We find that in both electron (${N>1}$) and hole doped (${N<1}$) cases, the electronic compressibility ${\kappa =  \frac{1}{N^2}\frac{{\rm d} N}{{\rm d} \mu}}$ diverges almost immediately when the chemical potential reaches the Hubbard bands.
Importantly, this divergence occurs not only when approaching the Hubbard bands from half filling (${\mu_2=-0.819}$\,eV and ${\mu_3=0.871}$\,eV), but also from the unfilled (${\mu_1=-1.068}$\,eV) and fully filled (${\mu_4=1.131}$\,eV) sides.
The divergence of the compressibility indicates the appearance of the PS region, which is additionally confirmed by the divergence of the charge susceptibility at the zero momentum ${{\bf q}=\Gamma}$.
This divergence appears when the static dielectric function in the charge channel $\epsilon^{\rm ch}_{\omega=0}({\bf q})$ approaches zero.
Importantly, the divergence of the compressibility and the charge susceptibility occurs at the same values of the chemical potential (Figure~\ref{fig:LE}), although these quantities are computed independently.
Extrapolating $\epsilon^{\rm ch}_{\omega=0}(\Gamma)$ to zero shows that the PS occurs at $19.5\pm1.5\%$ of doping.
At ${\mu_1<\mu<\mu_2}$ and ${\mu_3<\mu<\mu_4}$ the charge susceptibility remains divergent, defining the PS as regions of ``forbidden'' chemical potentials highlighted in red colour in Figure~\ref{fig:Phase_diagram}.

The identified PS is formed differently than in other Mott insulators~\cite{PhysRevLett.104.226402, PhysRevB.84.075161, PhysRevB.100.085104, PhysRevB.102.205127, PhysRevB.105.125140, PhysRevLett.130.066401, PhysRevB.110.L121109}, where the PS appears between the two connected branches of the $N(\mu)$ curve.
In our case, the $N(\mu)$ (red) curves in Figure~\ref{fig:Phase_diagram} exhibit an upward trend in the vicinity of all four PS boundaries.
This behaviour is linked to a strong asymmetry in the electronic density of states near the Fermi energy, which occurs when the chemical potential approaches the narrow Hubbard bands. 
This asymmetry causes a significant increase in the average electronic density due to strong charge fluctuations.
As we approach the PS by increasing $\mu$, which also increases $N$, we observe the expected rise in the $N(\mu)$ curve near $\mu_1$ and $\mu_3$. 
Conversely, when approaching the PS by decreasing $\mu$, the behaviour of the $N(\mu)$ curve near $\mu_2$ and $\mu_4$ becomes more complex.
Further away from the PS, where both the charge fluctuations and the asymmetry in the density of states are small, the electronic density initially decreases with decreasing $\mu$. 
However, in close proximity to the PS, the rapid increase in the strength of charge fluctuations and the asymmetry leads to a significant rise in $N$ even as $\mu$ decreases, resulting in a region of negative compressibility.
It is worth noting that since the charge dielectric function diverges within the PS regions, we cannot determine whether the two segments of the $N(\mu)$ curve on opposite sides of the PS regions are connected.

We also observe that the formation of the PS is related to a strong change in the electronic spectral function.
Figure~\ref{fig:A_dop} shows the momentum-resolved spectral function $A({\bf k}, E)$ plotted along the high-symmetry path in the Brillouin zone (BZ) that goes through the $\Gamma$, ${\text{M}=(\pi,\pi/\sqrt{3})}$ and K points.
Farther from the PS, at ${\mu=\mu_2+6}$\,meV (a) the spectral function corresponds to the usual case of a hole-doped Mott insulator~\cite{RevModPhys.68.13}, with the quasi-particle band at the Fermi energy (${E=0}$ depicted by the horizontal dashed white line) split from the lower Hubbard band (${E\simeq-0.2}$\,eV), which is reflected in the suppression of the spectral weight between these bands at ${E\simeq-0.1}$\,eV. 
Upon approaching the PS, the lower Hubbard band flattens, and at ${\mu=\mu_2+2}$\,meV (b) the $A({\bf k}, E)$ forms a pseudogap at ${E=0}$ at the incommensurate momenta, which transforms to a gap at the PS boundary determined by $\mu_2$ (c).
We also observe the development of incoherent bands below the Fermi energy in close proximity to the PS.
At ${\mu=\mu_2+2}$\,meV (b) these bands appear rather close to ${E=0}$, while at the PS boundary (c) they transform into almost straight lines localized at the incommensurate momenta where the gap is formed.  
The flattening of the Hubbard band and the emergence of incoherent bands below the Fermi energy stem from the increase in electronic density near the PS.

\begin{figure*}[t!]
\includegraphics[width=1\linewidth]{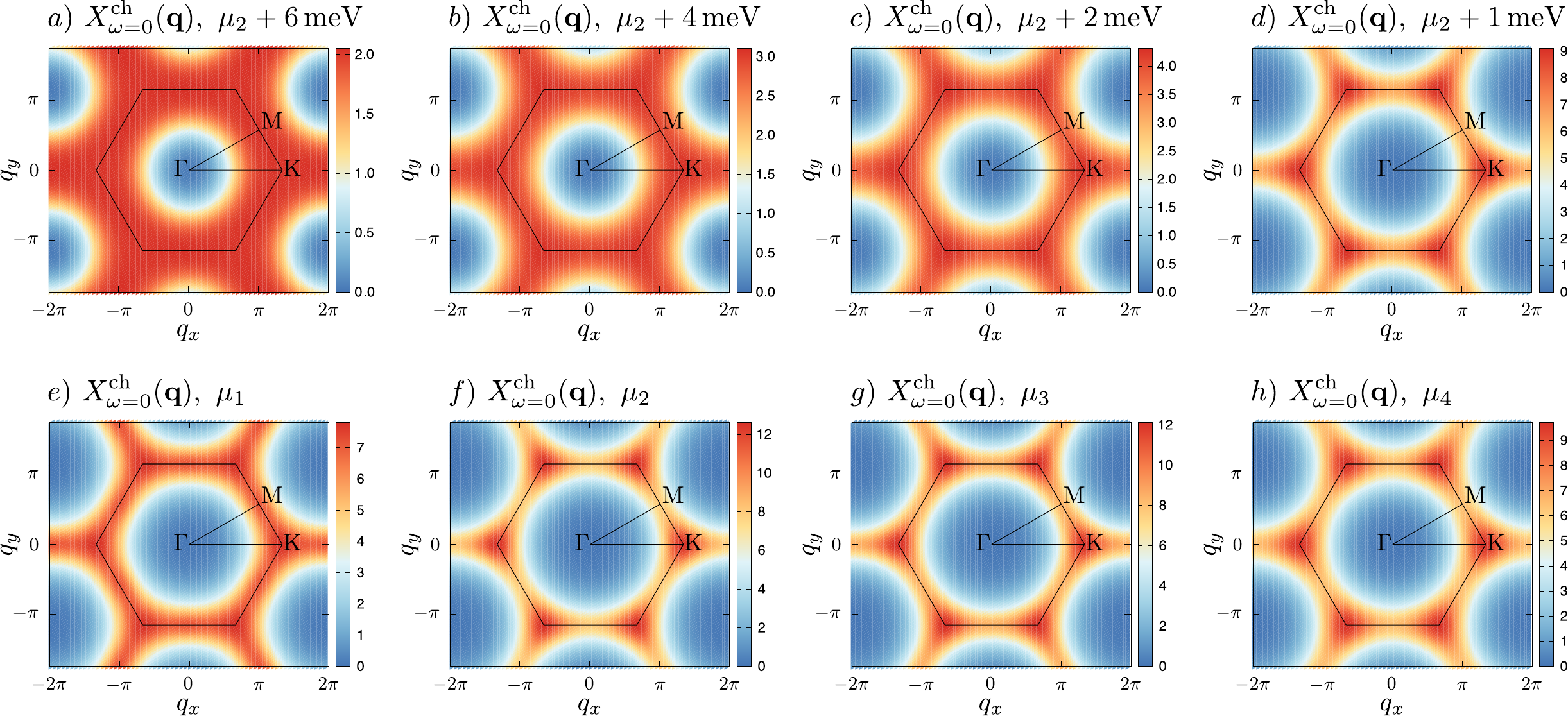}
\caption{Charge susceptibility. Top row: Evolution of the static charge susceptibility $X^{\rm ch}_{\omega=0}({\bf q})$ in the vicinity of the PS. The calculations are performed for ${\mu=\mu_2+6\,\text{meV}}$ (a), ${\mu_2+4\,\text{meV}}$ (b), ${\mu_2+2\,\text{meV}}$ (c) and ${\mu_2+1\,\text{meV}}$ (d). 
Bottom row: The static charge susceptibility obtained at different PS boundaries defined by the chemical potentials $\mu_1$ (e), $\mu_2$ (f), $\mu_3$ (g) and $\mu_4$ (h). 
All results are obtained at ${T=290\,\text{K}}$ and shown in the momentum space ${(q_x, q_y)}$. The first BZ is depicted by the black hexagon. The high-symmetry points $\Gamma$, K and M are labeled explicitly.
\label{fig:X}}
\end{figure*}

To understand the physical nature of the PS, let us examine the static charge susceptibility $X^{\rm ch}_{\omega=0}({\bf q})$ in the vicinity of $\mu_2$, as shown in the top row of Figure~\ref{fig:X}.
We find that the divergence of the dielectric function is not visible in the susceptibility due to a very small value of the polarization operator at the $\Gamma$ point. 
Instead, the largest value of the charge susceptibility corresponds to momenta at the edge of the BZ depicted by the black hexagon.
Far from the PS, at ${\mu=\mu_2+6}$\,meV (a) and ${\mu=\mu_2+4}$\,meV (b), where the largest charge dielectric function is respectively equal to ${\epsilon^{\rm ch}_{\omega=0}(\Gamma)=0.406}$ and ${\epsilon^{\rm ch}_{\omega=0}(\Gamma)=0.436}$, the charge susceptibility is relatively small and its maximum is distributed over a rather large part of the BZ.
As the PS is approached more closely (c), at ${\mu=\mu_2+2}$\,meV (${\epsilon^{\rm ch}_{\omega=0}(\Gamma)=0.156}$) the maximum of the susceptibility begins to localize at the edge of the BZ, but the value of the susceptibility still remains rather small.
Finally, in close proximity to the PS, at ${\mu=\mu_2+1}$\,meV, ${\epsilon^{\rm ch}_{\omega=0}(\Gamma)=0.007}$ (d) and $\mu_2$, ${\epsilon^{\rm ch}_{\omega=0}(\Gamma)=0.002}$ (f) the susceptibility drastically increases and displays a distinctive ``bow-tie'' shape with the maximum at the K point and the pinch-point at the M point.

The appearance of the bow-tie structure~\cite{PhysRevLett.87.047205, doi:10.1126/science.1064761, PhysRevLett.93.167204, PhysRevB.71.014424, doi:10.1126/science.1177582} in the spin susceptibility is one of the most direct probes of the spin ice state~\cite{balents2010spin}.
Spin ice is a highly frustrated magnetic state without long-range magnetic order. 
It lacks a conventional order parameter, which makes identifying this elusive state significantly more challenging.
In the case of a spin ice, the bow-tie structure in momentum space is a consequence of the dipolar correlations between the magnetic moments in real space, which decay as a power law ${\sim1/R^{D}}$ with the distance $R$, where $D$ is the dimension of the system~\cite{PhysRevLett.93.167204, PhysRevB.71.014424, doi:10.1126/science.1177582}.
This power-law form of correlations is reminiscent of the behavior of hydrogen bonding interactions in water and indicates the absence of static magnetic order, suggesting that spin correlations persist over long distances. 
In contrast, in magnetic systems with long-range order, spin correlations decay exponentially with distance because the system reaches a stable magnetic configuration. 
If the power law is reproduced exactly, the susceptibility at the M point shows a singularity. 
Deviation from the power law rounds this singularity, resulting in a rapidly decaying value of the susceptibility in the ${\text{M}-\Gamma}$ direction, which corresponds to a finite correlation length of the spin ice state~\cite{balents2010spin}. 
By analogy with this state, we identify the observed bow-tie form of the charge susceptibility with a charge ice state.
Remarkably, the bottom row of Figure~\ref{fig:X} demonstrates that this state is formed at all chemical potentials that define the boundaries of the PS regions and only in close proximity to the PS (${\epsilon^{\rm ch}_{\omega=0}(\Gamma)\lesssim0.01}$).
We find that the charge susceptibility does not exhibit a sharp singularity at the pinch-points (M). As shown in the top row of Figure~\ref{fig:X}, approaching the PS state (at $\mu_2$) enhances the peak at the M point and reduces its width. Since the present calculations capture only the precursor of the charge ice state, one can expect the charge susceptibility near the M point to become significantly sharper deeper inside the PS region, corresponding to the fully developed charge ice state.

\begin{figure}[b!]
\includegraphics[width=1\linewidth]{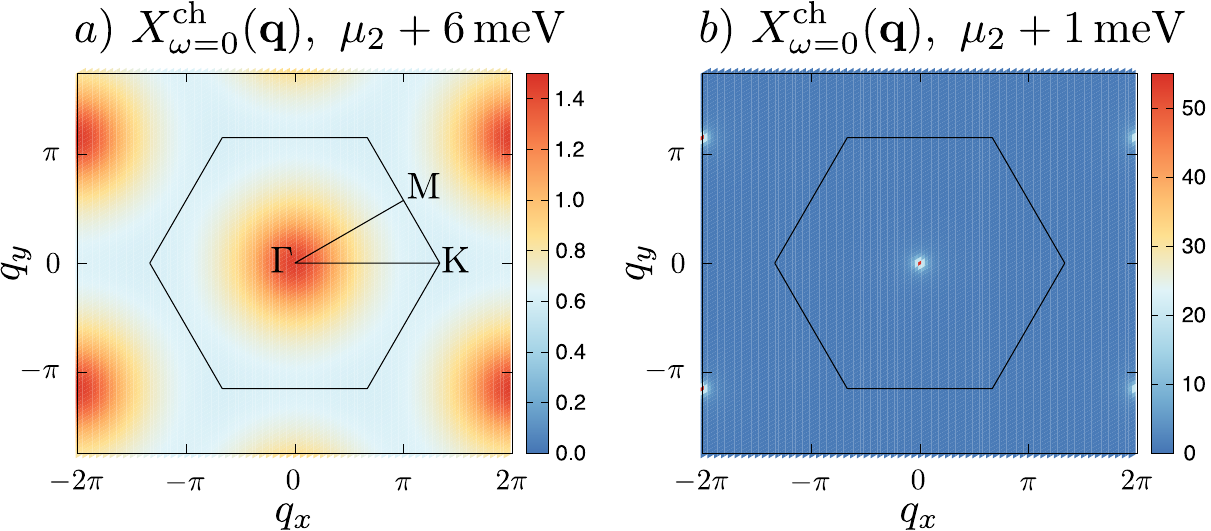}
\caption{Charge susceptibility in the absence of the long-range interaction. The static charge susceptibility $X^{\rm ch}_{\omega=0}({\bf q})$ calculated in the vicinity of the PS for ${\mu=\mu_2+6\,\text{meV}}$ (a) and ${\mu_2+1\,\text{meV}}$ (b) using \mbox{D-TRILEX} with only local interaction $U_0$.
\label{fig:Xch_U}}
\end{figure}

To clarify the role of the non-local Coulomb interaction in the formation of PS and the charge ice state, let us perform additional ``\mbox{D-TRILEX} ($U_0$)'' calculations, considering only the local part $U_0$ of the long-range Coulomb potential. 
The corresponding result for the average electronic density $N$ as a function of $\mu$ is shown in Fig.~\ref{fig:Phase_diagram} by the orange curves.
Within these new calculations, we find that PS is formed at the same values of the chemical potentials as in the \mbox{D-TRILEX} calculations with the full long-range Coulomb potential (red curve).
However, as shown in Fig.~\ref{fig:Xch_U}, when considering only the local Coulomb interaction, the charge susceptibility calculated in the vicinity of PS no longer displays a distinct bow-tie pattern in momentum space.
Instead, the charge susceptibility now displays the highest intensity at the $\Gamma$ point, which, in close proximity to PS, transforms into a delta-function-like peak [Fig.~\ref{fig:Xch_U}\,b)], indicating the formation of an ordinary PS state.
These results clearly demonstrate that PS is driven by the local interaction, while the non-local part of the Coulomb interaction is crucial for the formation of the charge ice state.

We have verified that the observed charge ice state is also present at lower temperatures.
At ${T = 145\,\text{K}}$, the PS state is found at approximately 8\% doping, and in its vicinity the charge susceptibility acquires a distinct bow-tie shape, qualitatively similar to that shown in the bottom row of Fig.~\ref{fig:X}.
This result is consistent with the finding of Ref.~\cite{vandelli2024doping}, which reports that the phase boundary of the charge instability observed in the doped triangular lattice Mott insulator Si(111):Pb shifts to lower doping levels as the temperature decreases.

\begin{figure}[t!]
\includegraphics[width=1\linewidth]{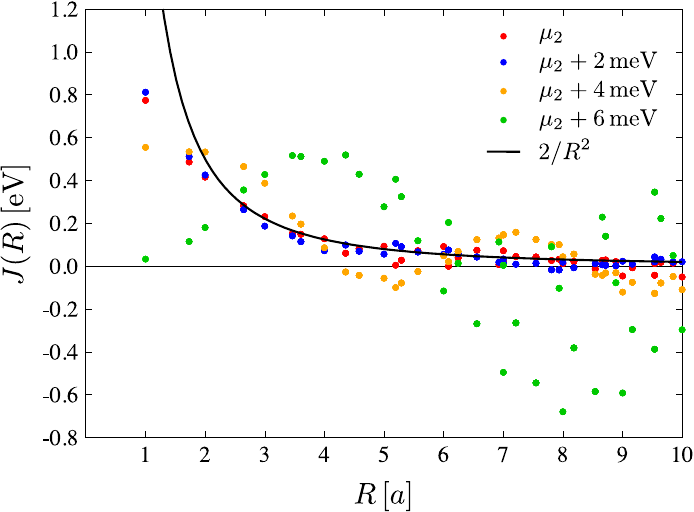}
\caption{Effective charge exchange interaction. The effective exchange interaction between the charge densities ${J(R)}$ calculated as a function of the real space distance $R$ in units of the lattice constant $a$. The results are obtained at ${T=290\,\text{K}}$ for ${\mu=\mu_2+6\,\text{meV}}$ (green), ${\mu_2+4\,\text{meV}}$ (orange), ${\mu_2+2\,\text{meV}}$ (blue) and ${\mu_2}$ (red). The fit function $2/R^2$ is depicted by the solid black line.   
\label{fig:J}}
\end{figure}

To gain more insights into this novel state, let us also explore the interaction between the electronic densities.
It has been shown in Ref.~\onlinecite{PhysRevB.99.115124} that the correlated electronic system in the vicinity of the charge instability can be mapped onto an effective Ising model
\begin{align}
H_{\rm eff} = \frac12\sum_{jj'}J_{jj'} \, \rho_{j} \, \rho_{j'},
\label{eq:Ising}
\end{align}
where ${\rho_{j}=n_{j}-N}$ is the difference of the electronic density on the site $j$ from the average occupation of the system.
The effective charge exchange interaction $J$, calculated following Ref.~\onlinecite{PhysRevB.99.115124} in the vicinity of the PS ($\mu_2$) as a function of the real space distance $R$, is shown in Figure~\ref{fig:J}. 
We find that further from the PS, at ${\mu=\mu_2+6}$\,meV (green) the charge exchange interaction is highly frustrated.
It has both, positive and negative values and the amplitude of $J(R)$ does not decay even at the distance of 10 lattice constants $a$. 
Reducing the chemical potential to ${\mu=\mu_2+4}$\,meV (orange) suppresses the frustration and the long-range tail of the charge exchange interaction.
Remarkably, only in a close proximity to the PS (blue and red), the long-range tail of the charge exchange interaction can be accurately fitted by the power law dependence of the distance ${J(R)\simeq2/R^2}$, as depicted by the black line.
This result is consistent with the behaviour of the charge susceptibility that reveals the bow-tie structure only at the PS boundary (Figure~\ref{fig:X}).
In addition, the deviation of the first few nearest-neighbor interactions $J$ from the power law may explain the ``rounded'' singularity at the M point of the obtained charge susceptibility.
We note that considering this power law form for the dipolar coupling between the magnetic moments was found important for modeling the spin ice behaviour in Ising pyrochlore magnets~\cite{PhysRevLett.83.1854, PhysRevLett.84.3430, doi:10.1126/science.1064761, PhysRevLett.101.037204, castelnovo2008magnetic,  jaubert2009signature, Bramwell_2020}. 
On the other hand, in some cases the power law correlations between the magnetic moments in real space can also be realized by considering only the nearest-neighbour exchange interaction~\cite{doi:10.1126/science.1064761}. 
Therefore, it is remarkable that the formation of the charge ice state is associated with the power law form of both the correlations between the charge densities, reflected in the bow-tie form of the charge susceptibility, and the effective charge exchange interaction.

\section{Conclusion}

In this work, we have investigated collective electronic instabilities in the monolayer Nb$_3$Cl$_8$. 
We have found that at half filling, the considered material lies deep in the Mott insulating phase and surprisingly does not reveal any tendency towards the formation of a charge or spin ordered state, contrary to some other Mott insulators with triangular lattice geometry~\cite{PhysRevLett.120.196402, vandelli2024doping}.
Upon doping, the system exhibits a region of PS, detected by a simultaneous divergence of the charge susceptibility (or dielectric function) at zero momentum $\Gamma$ and the electronic compressibility.
The critical doping required for the formation of the PS can be estimated through the extrapolation of the dielectric function to zero, resulting in an approximate value of 20\% doping at ${T=290\,\text{K}}$ and 8\% doping at ${T=145\,\text{K}}$.
This suggests that PS in monolayer Nb$_3$Cl$_8$ could potentially be realized experimentally, given the achievable doping levels in other two-dimensional systems, such as high-temperature superconducting cuprates.

We observe that away from the PS, the charge susceptibility is rather small and weakly momentum-dependent. 
In turn, the effective exchange interaction between the charge densities is highly frustrated and long-ranged. 
However, in close proximity to the PS, the charge susceptibility dramatically increases and reveals a distinctive bow-tie pattern in momentum space, reminiscent of the magnetic susceptibility observed in the spin ice state.
Simultaneously, the form of the charge exchange interaction undergoes a drastic change and acquires a power law dependence in real space.

Interestingly, the charge ice state is observed only when accounting for the effect of the full long-range Coulomb potential, while with only local interactions, the system reveals an ordinary PS state. 
This can be explained by the fact that geometrical frustration likely plays an important role in the development of the charge ice state, similarly to the spin ice state, as it prevents the formation of long-range order.
Incorporating the long-range Coulomb interaction enhances the frustration inherent in the triangular lattice, which, apparently, is not sufficient for the formation of the charge ice state if only the local interaction is considered. 
In this context, the single-band model for Nb$_3$Cl$_8$ provides a unique platform for exploring fascinating many-body effects driven by frustration.

We find that the development of the charge ice state is accompanied by the formation of a gap and incoherent bands in the electronic spectral function.
The appearance of the latter below the Fermi energy, along with the flattening of the Hubbard band, reflects the mechanism through which the electronic density increases upon approaching the PS.
In turn, the gap in the electronic spectrum should not be seen as an indicator of an insulating phase, because the gap occurs at the incommensurate doping of the system. 
Instead, we relate it to the PS, which emerges as a consequence of the formation of the charge ice state.
Indeed, unlike a conventional Ising model for spins, the introduced effective Ising model~\eqref{eq:Ising} has an additional constraint ${\sum_{i}\rho_i=0}$ that is needed to fix the total occupation of the system.
Unfortunately, we cannot access the real space structure of the charge ice state because we cannot perform symmetry-broken calculations inside the PS region.
However, as in artificial spin ices~\cite{wang2006artificial, RevModPhys.85.1473}, there one could expect the formation of ``ferromagnetic'' domains with higher and lower occupation than the averaged one, hence the PS.

Therefore, we find that the identified charge ice state, aside from the aforementioned constraint, is surprisingly similar to the spin ice state. 
It is also striking that the charge ice state has not been found in other Mott insulators with similar properties. 
For example, a system of Pb adatoms disposed periodically on a Si(111) surface also exhibits a triangular lattice geometry, features a strong and long-ranged Coulomb interaction, and lies deep in the Mott insulating state. 
Instead, upon doping, it reveals a charge density wave instability corresponding to the divergence of the charge susceptibility at a finite wave vector~\cite{vandelli2024doping}. 
This emphasizes the necessity for a deeper understanding of the mechanism responsible for the formation of the charge ice state.

\begin{acknowledgements}
The author thanks Alexander Kowalski and Giorgio Sangiovanni for the guidance in the \textsc{w2dynamics} calculations. 
The author also thanks Malte R{\"o}sner, Mikhail Katsnelson, Alexander Lichtenstein, Leon Balents, Maria Chatzieleftheriou and Luca de' Medici for the fruitful discussions and valuable comments. 
The help of the CPHT computer support team is acknowledged as well.
\end{acknowledgements}

\bibliography{Ref}

\end{document}